%% file: main.tex
\documentclass[conference]{IEEEtran}
\IEEEoverridecommandlockouts
\usepackage{algorithm}
\usepackage{algpseudocode}
\usepackage{cite,url,bm}
\usepackage{amsmath,amssymb,amsfonts}
\usepackage{graphicx}
\usepackage{textcomp}
\usepackage{xcolor,soul}
\usepackage{authblk}
\usepackage{caption}
\usepackage{subcaption}
\usepackage{enumitem}
\usepackage{tikz}
\usepackage{flushend}
\usepackage{booktabs}
\usepackage{array}
\usepackage{float}
\usepackage{longtable}
\usepackage{array}
\usepackage{multicol}
\usepackage{multirow}
\usepackage{lscape}
\def\BibTeX{{\rm B\kern-.05em{\sc i\kern-.025em b}\kern-.08em
    T\kern-.1667em\lower.7ex\hbox{E}\kern-.125emX}}

\title{An Analysis of  Resource Allocation  and User Association Strategies in Space-Air-Ground Integrated Networks}
\author[1]{Siri Vennela Geddam}
\author[2]{Sruthi Ilapuram}
\author[2]{Kamesh Namuduri}
\author[1]{K L V Sai Prakash Sakuru}
\affil[1]{National Institute of Technology, Warangal, Telangana, India}
\affil[2]{Department of Electrical Engineering, University of North Texas, Denton USA}

\begin{document}
\maketitle
\pagenumbering{arabic}
\setcounter{page}{1}
\input{abstract}
\input{introduction}

\input{literature}

\input{systemmodel}
\input{userAssociation}
\input{mobility}

\input{resourceAllocation}
\input{simulations}
\input{conclusions}

\input{ack}
\bibliographystyle{unsrt}
\bibliography{ref}
\end{document}

%% file: abstract.tex
\begin{abstract}

Space-Air-Ground-Integrated Networks (SAGIN) enable seamless data connectivity for applications such as smart transport, healthcare, smart cities, and disaster response through the coordinated use of low-earth orbit (LEO) satellites, base stations mounted with uncrewed aerial vehicles (UAV), and terrestrial infrastructure. This paper provides a detailed analysis of resource management frameworks, reviews the literature, and evaluates key methods such as alternating optimization (AO), damped iterative water filling (DIWF), and genetic algorithms (GA) for resource allocation. MATLAB simulation results benchmark these algorithms across 10,000 trials, demonstrating robust, fair, and low-latency resource allocation.  In addition, this paper also analyzes strategies for user association with terrestrial and aerial base stations during emergencies and network overloads. The main contributions include a comparative assessment of resource allocation strategies in SAGIN and an in-depth analysis of user association policies for emergency scenarios. The study provides guidance for designing resilient and efficient next-generation networks. Potential future research directions include investigating satellite handover and multi-domain orchestration for SAGIN deployments.

\end{abstract}

%% file: introduction.tex
\section{Introduction}
Uninterrupted and robust network connectivity is essential for real-time applications such as autonomous navigation, disaster recovery, and military operations \cite{zhang2025uav,Navin2021}. However, these demands cannot be fully met by conventional terrestrial networks. Fortunately, space-air-ground integrated networks (SAGINs) provide a promising solution to this challenge. SAGIN is an emerging network architecture that refers to the vertical integration of the low-earth orbit/medium-earth orbit/geostationary-earth orbit (LEO/MEO/GEO) satellites in space, high-altitude platforms/uncrewed aerial vehicles (HAPs/UAVs) in air, as well as terrestrial base stations (TBSs) and other networks on the ground. SAGIN has the ability to provide ubiquitous coverage to underserved and isolated areas by maintaining a flexible, resilient, and stable networking architecture that allows the offloading of traffic from congested terrestrial networks to one of the other aforementioned platforms, which is selected depending on the specific conditions of the scenario under study. The 3rd Generation Partnership Project (3GPP) beginning with release 17 and continuing through release 18 and beyond has shown that non-terrestrial networks (NTNs) and SAGIN are necessary to achieve the global 5G performance criteria of ITU IMT-2020 \cite{Navin2021}. As such, the implementation of SAGIN is considered essential in order to see next generation 5G/6G strategies go from the conceptual stage to full-scale deployment. Although there are many use-cases across the spectrum of modern industry where SAGIN proves beneficial, public safety and disaster response have been selected as the focus points in this paper to illustrate the benefits that SAGIN provides.

\subsection{Context: Emergency Communications} 
Traditional terrestrial network infrastructure is prone to being damaged in emergency scenarios such as hurricanes and earthquakes; however, SAGIN can rectify this by providing immediate communication between the affected population and the rescue crew. Even in non-emergency scenarios such as major political events or sports events where massive groups of people are congregated together, users experience a decline in quality of service (QoS) due to the limited capacity of terrestrial base stations. In both scenarios, SAGIN helps to offload the call traffic from TBSs to base stations mounted on UAVs, referred to as aerial base stations (ABSs). Satellites provide the necessary backhaul link between UAVs and the cellular network on the ground. From an implementation perspective, there are two main problems that must be addressed in order to properly utilize SAGIN: user association and resource allocation.

\subsection{User Association / Offloading}
In SAGIN, user association is the process of linking a user to a particular base station located either in the air (ABS) or on the ground (TBS). User mobility, node density, resource availability, node capacity, signal-to-interference plus noise ratio (SINR) at the user are the relevant parameters taken into account for associating users with ABS or TBS. In emergency scenarios, user association refers to connecting users to a particular base station based on a predetermined criterion such as proximity or priority. In general, user association refers to offloading of users from one base station to another, depending on availability of resources.

\subsection{Resource Allocation}
The objective of resource allocation in SAGIN is to increase throughput, reduce latency, ensure fairness among users, distribute the load uniformly on the network, and minimize energy. The resources of concern here include bandwidth, power, computing, as well as the network at the ground, air, and space layers. Due to the dynamic topology, heterogeneity, and stringent QoS requirements of the network, resource allocation in SAGIN is a crucial and challenging issue.  

\subsection{Major Contributions}
The key contributions of this paper are summarized as follows:

\begin{itemize}
    \item \textbf{Comprehensive survey and categorization} of recent SAGIN research, covering resource allocation, user association, UAV placement, and task offloading.  
    
    \item \textbf{Modeling of representative deployment scenarios} (single-TBS failure, multi-TBS outage, and large-scale user offloading) to illustrate the resilience benefits of SAGIN.  
    \item \textbf{Formulation of an orthogonal frequency division multiple access (OFDMA)-based resource allocation problem} for ABS-assisted SAGIN, incorporating subcarrier exclusivity, power budget and QoS constraints, posed as a mixed integer nonlinear programming (MINLP) problem.  
    
    \item \textbf{Comparative performance evaluation of algorithms} -- alternating optimization (AO), damped iterative water filling (DIWF) and genetic algorithm (GA) -- using extensive monte carlo (MC) simulations.  
    
    \item \textbf{Integrated analysis of resource allocation and user association strategies}, including SINR, distance, and priority-based policies, with insights on performance, fairness, convergence, and robustness for next-generation SAGIN deployments.  
\end{itemize}

\subsection{Organization}
The remainder of the paper is organized as follows. Section \ref{sect:lit} reviews the related literature, while sections \ref{system architecture}-\ref{mobility} present the SAGIN system architecture, spectrum considerations, user association problem, and mobility aspects. Section \ref{resource allocation} formulates the resource allocation problem and introduces the solution algorithms, followed by the simulation setup and the results in Section \ref{simulations}. Finally, Section \ref{concl} concludes the paper with key findings, helpful insights, and future research directions. 

\subsection{Scope of this Research}
This paper focuses on a simplified SAGIN configuration consisting of one LEO satellite in the space layer, multiple ABSs in the air layer, and several TBSs in the ground layer. The analysis is limited to user association between ABS and TBS and resource allocation under OFDMA-based communication. Satellite handover, multi-satellite constellations, cross-domain orchestration, and UAV trajectory optimization are beyond the scope of this work. These aspects, along with topics such as real-time traffic dynamics, mobility-aware resource scheduling, and security/privacy considerations, will be investigated in subsequent works. Therefore, the present study serves as a foundation, providing tractable models and comparative evaluations that can be extended to more complex SAGIN deployments.

\begin{table}[h]
\centering
\caption{Main nomenclature}
\label{Table-Nomenclature}
\begin{tabular}{|c|p{5cm}|}
\hline
\textbf{Word} & \textbf{Description} \\
\hline
SAGIN & Space-air-ground integrated network \\
\hline
LEO & Low Earth orbit \\
\hline
UAV & Uncrewed aerial vehicle \\
\hline
HAP & High-altitude platform \\
\hline
MEO & Medium Earth orbit \\
\hline
GEO & Geostationary Earth orbit \\
\hline
TBS & Terrestrial base station \\
\hline
3GPP & 3rd generation partnership project \\
\hline
4G & Fourth generation \\
\hline
5G & Fifth generation \\
\hline
6G & Sixth generation \\
\hline
NTN & Non-terrestrial network \\
\hline
ABS & Aerial base station \\
\hline
SINR & Signal-to-interference-plus-noise ratio \\
\hline
QoS & Quality of service \\
\hline
SCA & Successive convex approximation \\
\hline
DQN & Deep Q-network \\
\hline
PODMAI & Partially observable deep multi-agent active inference \\
\hline
DRL & Deep reinforcement learning \\
\hline
LYMOC & Lyapunov mixed integer linear programming based optimal cost \\ 
\hline
DL & Deep learning \\
\hline
RL & Reinforcement learning \\
\hline
AO & Alternating optimization \\
\hline
GA & Genetic algorithm \\
\hline
WF & Water filling \\
\hline
DIWF & Damped iterative water filling \\
\hline
CNN & Convolutional neural network \\
\hline
B-CNN & Binary CNN \\
\hline
SNR & Signal-to-noise ratio \\
\hline
NOMA & Non-orthogonal multiple access \\
\hline
LTE & Long term evolution \\
\hline
LoS & Line-of-sight \\
\hline
MIMO & Multiple-input multiple-output \\
\hline
MT & Maximum throughput \\
\hline
PF & Proportional fairness \\
\hline
mmWave & Millimeter wave \\
\hline
SDN & Software-defined networking \\
\hline
NFV & Network function virtualization \\
\hline
D2D & Device-to-device \\
\hline
STIN & Satellite-terrestrial integrated network \\
\hline
AI & Artificial intelligence \\
\hline
COW & Cell on wheels \\
\hline
CBRS & Citizens broadband radio service \\
\hline
OFDMA & Orthogonal frequency division multiple access \\
\hline
MC & Monte carlo \\
\hline
ADMM & Alternating direction method of multipliers \\
\hline
MINLP & Mixed integer non-linear programming  \\
\hline
\end{tabular}
\end{table}


%% file: literature.tex
\section{Literature}
\label{sect:lit}
In the existing literature, SAGIN has been proposed as a solution for a variety of use cases that include extending coverage area on the ground, user offloading from TBS to ABS, mobile edge computing, service caching, and network slicing services, among others. In each of these scenarios, resource allocation has been identified as the most critical research problem, followed by user association within the ground, air, and space layers of the SAGIN. Measurable outcomes discussed in the existing literature include improvements in service quality, resource accessibility, throughput, sum transmission rate, reliability, latency, and power consumption. This section is dedicated to further investigation into these use case scenarios, research topics, and performance metrics, in order to illustrate a wholesome picture of the existing SAGIN-related literature. 

\begin{figure}[htbp]
  \centering
 \begin{subfigure}{0.55\textwidth} \includegraphics[width=0.75\textwidth]{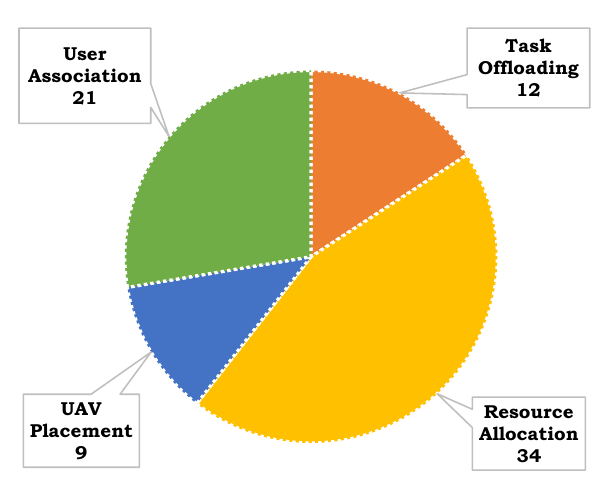}\centering \caption{Research Problems} \label{fig:presearchproblem} \end{subfigure} \begin{subfigure}{0.55\textwidth}\centering \includegraphics[width=0.60\textwidth]{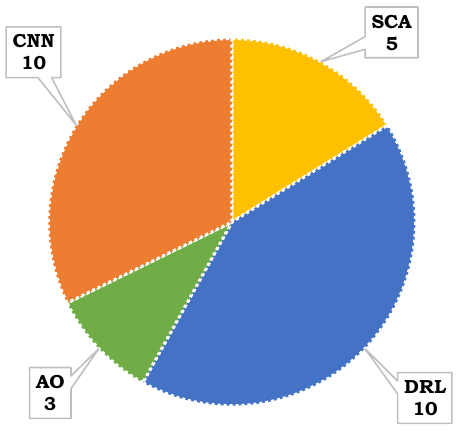} \caption{Algorithms} \label{fig:algorithms} \end{subfigure} \begin{subfigure}{0.55\textwidth}\centering \includegraphics[width=0.75\textwidth]{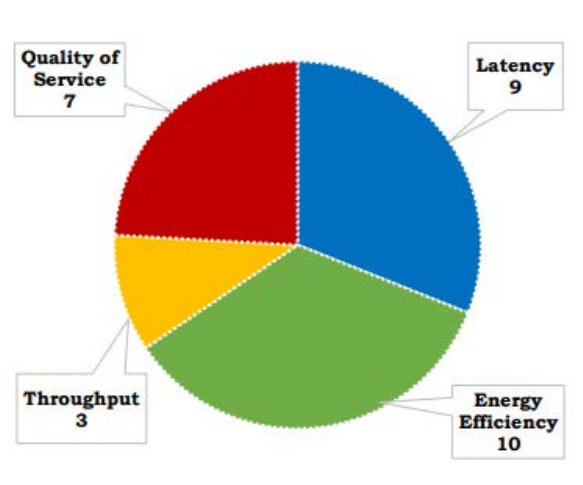} \caption{Performance Metrics} \label{fig:metrics} \end{subfigure}
  \caption{ An overview of current literature in SAGIN in terms of research problems, algorithms, and performance metrics, based on a sample study of 50 papers.}
  \label{fig:sagin_overview}
\end{figure}

\subsection{Overview of literature}
This section provides an overview of the state-of-the-art in SAGIN in terms of three aspects: 1) research problems of significant interest, 2)  algorithms developed, and 3) performance metrics used in simulation studies. The pie charts shown in Figure \ref{fig:sagin_overview} provide such an overview. 

\paragraph{Research Problems} In terms of research problems, resource allocation constitutes the largest share of the lot comprising 34 of 50 total papers, thus highlighting its fundamental role in managing limited spectrum and power resources. User association and task offloading are also widely studied in 21 of the 50 papers, reflecting the growing importance of dynamic connectivity and the distribution of computing resources in heterogeneous networks. In addition, the UAV placement problem has only been studied in 9 papers, indicating that this is a research area requiring more attention. It should be noted that this distribution corresponds specifically to the sample study conducted here and does not reflect the overall distribution of the existing literature; however, it does act as a measure of the relevance of key research topics related to SAGINs.

\paragraph{Algorithms} With respect to algorithmic approaches, the literature shows a balance between optimization-based and learning-driven methods. Successive convex approximation (SCA), convolutional neural networks (CNN) and deep reinforcement learning (DRL) are the most frequently applied algorithms, demonstrating a trend toward hybrid frameworks that combine mathematical rigor with adaptive intelligence. Meanwhile, AO is applied less frequently, indicating a slow transition from conventional iterative methods.

\paragraph{Performance Evaluation} The performance metrics that existing studies currently emphasize are energy efficiency and latency, which highlights the interest in current research in sustainability and high-speed connectivity. In contrast, throughput is the least emphasized performance metric, due to its indirect consideration in other related performance metrics such as quality of service.

\subsection{Use Case Scenarios}
One of the important use cases of SAGIN is during emergency scenarios. For example, immediately after a natural disaster, a SAGIN supported by a satellite backhaul can be deployed to substitute for a TBS that might have become dysfunctional.  A SAGIN can also be deployed to expand network coverage areas and improve service continuity in situations where terrestrial network infrastructure is inadequate or compromised. Thus, the two key use cases that highlight the utility provided by SAGINs are: 1) supplementing existing networks to facilitate deployment of emergency communications, and 2) offloading traffic of congested networks during dense public gatherings. Among the practical benefits of SAGINs are rapid deployment times, dynamic repositioning capabilities, and an increased ability to maintain line-of-sight (LoS) connectivity, all of which are especially valuable in the wake of a disaster when terrestrial networks have been rendered inoperable. Additionally, SAGINs also prove useful during large public events where user density temporarily exceeds the capacity of the fixed infrastructure. In such cases, ABSs can remedy the situation by dynamically offloading traffic and optimizing spectrum utilization between network nodes.  A SAGIN with beam-hopping LEO satellites and dynamic user association algorithms is another strategy that is helpful during emergency scenarios \cite{onyekwelu2024extended}. This technique is good at providing scalable connectivity during transient mass gatherings. 

Edge computing scenarios highlight the importance of effective resource management. A SAGIN can act as a mobile edge server in a deep Q-network (DQN)-based task offloading system for multi-access edge computing, allowing for effective and dynamic resource allocation under highly variable network conditions \cite{zhai2025joint}. They can offload traffic from overloaded devices during disaster management operations and live events, reduce overall latency in the network, and strategically adjust to task dependencies. Therefore, the deployment of SAGIN is useful to ensure reliable, flexible, and on-demand communication services. SAGINss are an essential component of next-generation wireless communication infrastructure, whether to restore network operation following a disaster or to meet a higher demands from users in real time.

\subsection{User Association / Offloading}

The user association in SAGIN is crucial for maintaining optimal connectivity and resource usage across network layers.  When a user is close to a TBS and has relatively good LoS resulting in a strong signal strength, they are typically connected to that TBS. This process is called user association. The user association is determined by the distance from the user to the TBS as well as SINR. Furthermore, when a good LoS to the TBS cannot be achieved, signal deterioration occurs and SINR becomes the primary determining factor for user association \cite{onyekwelu2024extended,xu2024joint}. In edge computing applications, distance-based limits ensure that task offloading is limited to edge servers within the coverage area \cite{zhai2025joint,xu2024joint}. 

\paragraph*{Modeling user association} Channel gain, defined using standard distance-based path loss models, and SINR, defined in terms of transmit power, noise, interference, and distance-dependent gain \cite{ding2025novel}, \cite{vandrl}, \cite{li2025joint} and \cite{peng2025intelligent} are typically used to determine user association. These models are also used to determine feasible relationships for UAVs and satellite connectivity. In satellite networks, transmission rates based on the signal-to-noise ratio (SNR), and link quality influence access selection decisions \cite{peng2025intelligent}, \cite{triyanto2025computation} and task offloading behavior. Some techniques, particularly those that use DRL, encode distance-based data, such as hop count, as well as system resources, into feature vectors for softmax-based node selection, with a focus on proximity and capacity \cite{chen2025joint}.

In other models, binary association variables are also used to represent the selection of a base station (TBS, HAP, or satellite) for a user, which is limited by signal quality and distance-based channel gain \cite{chen2025vehicular}. These models explicitly optimize user association by taking path loss, interference, and available resources into account, with the goal of maximizing throughput, while maintaining fairness \cite{wang2021incorporating}, \cite{vcauvsevic20214g}, \cite{shah2020comparison}, \cite{sheng2023coverage} and \cite{matracia2022post}. 

In general, the literature implies that effective user association techniques in SAGIN require a hybrid strategy that combines distance-aware and SINR-based decisions to adapt to dynamic situations such as user mobility, environmental factors, and varying link quality.

\subsection{Resource Allocation}

Resource allocation problems have been extensively studied in the literature. A survey of resource allocation methods \cite{onyekwelu2024extended,ding2025novel,liang2024resource} highlights the challenges and opportunities in SAGIN.

\paragraph*{Frameworks for resource allocation} Recent research highlights intelligent, adaptive, and integrated resource management frameworks that simultaneously optimize communication, storage, computation, and energy resources while meeting stringent constraints on delay, energy efficiency, and fairness. These methods combine advanced artificial intelligence techniques such as DRL, multi-agent systems, and federated learning with advanced optimization techniques such as convex and non-convex programming, mixed-integer programming, Lyapunov-based dynamic methods, and heuristic/ metaheuristic algorithms \cite{matracia2022post,alqurashi2022maritime, wang2021incorporating,liu2025satellite}. In applications such as emergency UAV-assisted networks, maritime communications, vehicular edge computing, and Ocean of Things, these hybrid approaches allow dynamic task offloading, bandwidth scheduling, power control, UAV/satellite trajectory optimization, and joint caching strategies \cite{zhang2022distributed, ding2025novel, vandrl}. Frameworks that use actor-critic architectures, Markov decision processes, and knowledge-driven techniques show notable improvements in throughput, energy efficiency, latency reduction, fairness, and network lifetime, along with significant gains in overall adaptability and robustness \cite{huang2023delay, nguyen2022computation, alsharoa2020improvement}.

\paragraph*{Resource allocation in next generation wireless networks} Resource allocation has become a key topic in next generation wireless networks \cite{degambur2021resource, hao2021investigation}. In order to balance throughput, fairness, and delay, the allocation of 4G long-term evolution (LTE) resources focused on the management of radio resources at the base station using scheduling algorithms such as maximum throughput (MT) and proportional fairness (PF) \cite{degambur2021resource}. With the advent of 5G, resource allocation became multidimensional. Millimeter wave (mmWave) spectrum, beamforming, non-orthogonal multiple access (NOMA), massive multiple-input multiple-output (MIMO) and dynamic spectrum sharing \cite{salah2021comparative, vcauvsevic20214g, vishwakarma2024comparative} necessitated the allocation of not only spectrum but also computing and storage, which was made possible by the implementation of software-defined networking (SDN) / network function virtualization (NFV) and network slicing \cite{shah2020comparison, degambur2021resource}. 

\paragraph*{Resource allocation in complex scenarios} Small cells, device-to-device (D2D) communications, and dual connectivity in heterogeneous 5G networks made resource management even more difficult by necessitating coordinated, often intelligent policies for spectrum, load balancing, and power \cite{xu2021survey,agiwal2021survey}. Resource allocation has been extended to the radio, computational and network levels by enabling 5G and the upcoming 6G technologies such as massive MIMO, cognitive radio, programmable networks, and network slicing \cite{shah2020comparison, salah2021comparative}. The transition to 6G, particularly with the satellite-terrestrial-integrated network (STIN), requires multidomain resource scheduling, where adaptive frameworks driven by Artificial Intelligence (AI) enable dynamic management of spectrum, energy and computation across terrestrial, aerial, and space assets \cite{sheng2023coverage, saeed2021point, yan2020interference}.

\paragraph*{Machine-learning based resource allocation}


 Machine learning methods such as CNN are increasingly being used to optimize resources \cite{zhang2019convolutional,goswami2021neural,zhang2020application,li2022convolutional,rahman2020convolutional,jia2020intelligent,lei2019learning}. Binary CNN are suggested for dynamic resource management \cite{qi2024application}. CNN-based approaches are also proposed for deep power control \cite{lee2018deep} at reduced computational complexity. Evolution of cellular networks, QoS requirements, and features for supporting multimedia services have been discussed in \cite{hajlaoui20204g,mihret20214g,davronbekov2021features}.

\paragraph*{Criteria for resource optimization} Criteria such as spectrum efficiency, latency, fairness, QoS, energy consumption, and computing utilization are used to evaluate the effectiveness of resource allocation techniques \cite{degambur2021resource, sulistiyo2023signal}. Designing intelligent and scalable algorithms for ultra-dense, heterogeneous, and integrated networks, ensuring smooth hand-off, protecting privacy and security, and striking a balance between sustainability and efficiency remain challenges despite advancements \cite{xu2021survey, yan2020interference, saeed2021point}.  In order to achieve the goals of 5G, 6G, and beyond, resource allocation is therefore shifting from conventional, static radio scheduling to a holistic, adaptive, and intelligent multi-resource orchestration paradigm.


%% file: systemModel.tex
\section{SAGIN System Architecture and Deployment Scenarios}
\label{system architecture}

\begin{figure}[h]
    \centering    
    \includegraphics[width=2.5in]{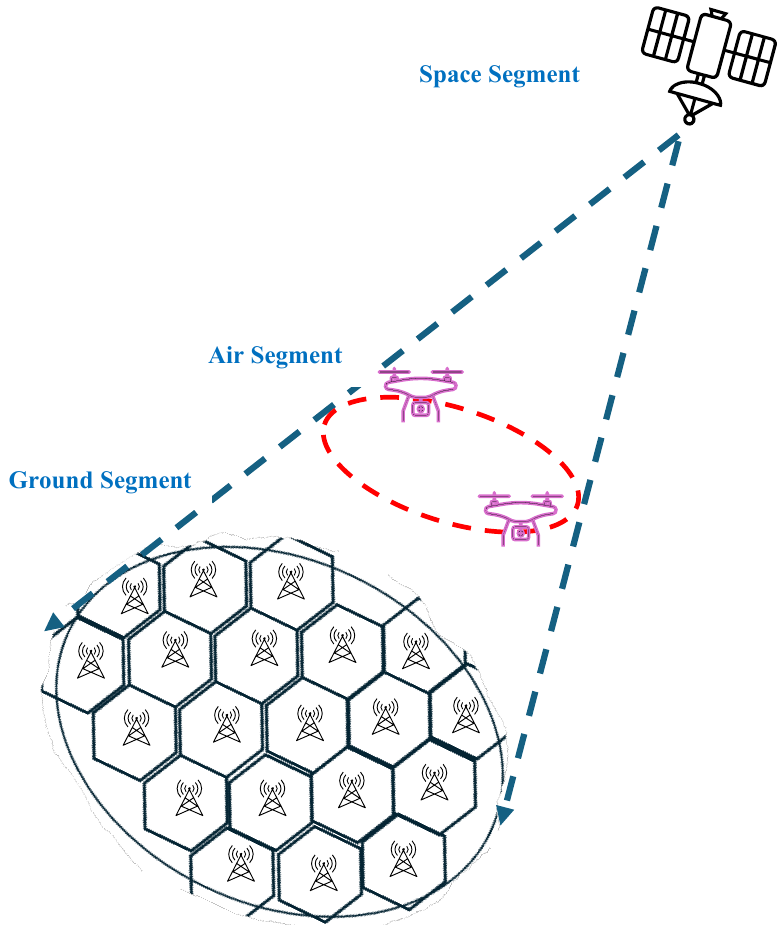}
     \caption{A SAGIN architecture consisting of space, air, and ground segments}
    \label{fig:SAGIN}
\end{figure}

Wireless networks have to ensure continuous user connectivity even during times when one or more TBSs are down due to natural or man-made disasters. A SAGIN can be seen as a vertical and intelligent network solution that provides resiliency to TBS failures. This section describes the system architecture for a SAGIN and presents a few deployment scenarios. A SAGIN architecture with one LEO satellite, two UAVs, and several TBSs is illustrated in Figure \ref{fig:SAGIN}. 

\subsection{Scenario 1: Service continuity when single TBS is down}
When a TBS in a cell stops working because of a repair or a natural disaster, several ways are used to keep users connected. Cell breathing is the process by which nearby TBSs dynamically modify their antenna tilt and transmission power to increase coverage in the affected area. Depending on acceptable SINR levels and available resources, users within the outage zone are reallocated to the closest TBS or an available airborne platform. An ABS mounted on a UAV can be used to temporarily cover areas if ground-based fallback options are inadequate. In addition, satellite links could be used to keep the ABS and the core network connected in the event that the terrestrial backhaul is damaged. When combined, these strategies help close the average gap and guarantee temporary service continuity until the damaged base station is fixed. 
\subsection{Scenario 2: Service continuity when multiple TBSs in a cluster are down}
To guarantee service continuity in the event of a widespread outage that impacts several base stations in a cluster, for example, during natural disasters or coordinated attacks, a multi-tiered response is put in place. To ensure smooth geographical coverage, a number of UAVs are sent out to cover the impacted cells, either by establishing a hovering network or by flying along pre-arranged flight paths. Wide-area resilience is ensured by LEO satellites, which serve as backup relay nodes or offer backhaul communication for UAVs. In addition, temporary restoration of terrestrial connectivity can be achieved by using quickly deployable mobile ground stations, including cell on wheels (COW). In more extreme situations, D2D communication may be used by local user equipment to create ad hoc mesh networks to support basic services. The SAGIN framework uses an SDN enabled control plane to manage user associations with the most dependable nodes and dynamically distribute network resources. This adaptable and hierarchical approach reduces service interruption and increases the efficiency of disaster recovery operations.
\subsection{Scenario 3: User offloading from TBSs to ABSs during public events}
During major public events, such as concerts, sporting events, or festivals, TBSs often experience congestion due to a sudden increase in user density. Users are offloaded from TBSs to ABSs mounted on UAVs in order to preserve the intended QoS. These communication- capable UAVs are either launched in real-time or pre-positioned to hover over densely populated regions. Offloading decisions are governed by dynamic load balancing algorithms that take into account SINR, user density, and the available capacity of each node. Signal quality is evaluated using path loss models, and users switch to an ABS when it offers a high SINR or when the TBS exceeds its load threshold. UAVs use satellite relays, mmWave, or microwave technology to maintain high-throughput backhaul links to the core network. Standard-compliant handover processes, such as x2 or NG handovers in 5G, are used to handle the switch from TBS to ABS with the least amount of disturbance possible. When combined, these approaches increase network capacity, avoid congestion, and further improve user experience in situations of high demand.

\section{Spectrum Considerations in SAGIN}
\label{Spectrum considerations}

This subsection examines the use of frequency bands for space, air, and ground segments.

\paragraph{Space Segment} The purpose of the space segment is to provide backhaul connectivity. The Ku band (12–18 GHz) and the Ka band (27–40 GHz) are the main frequency bands used by the LEO satellites. In the Ku band, the 10.7–12.7 GHz region is frequently used for user downlink (from satellite to ground). Similarly, in the Ka band, the 27.5-29.1 GHz and 14.0-14.5 GHz regions are mainly utilized for user uplink (from the ground to the satellite). This Ka band is also occasionally used for user downlink. These high frequency bands are more vulnerable to weather and rain fade, but they also have benefits in terms of data capacity, bandwidth, and antenna size. Due to reduced susceptibility to weather interference, the L band is also utilized for certain LEO applications, especially in maritime and long-distance aircraft communications. 

\paragraph{Air Segment}
The 150 MHz wide band situated between 3.55 and 3.7 GHz, referred to as the citizens broadband radio service (CBRS), is commonly used for private mobile networks. In this research, the CBRS band is assumed for base stations mounted on UAVs. Since the CBRS spectrum is shared, a variety of users, including incumbent users, priority license holders, and users with general access use this spectrum. 

\paragraph{Ground Segment}
5G networks function in three primary frequency bands: low band, mid band, and high band (referred to as mmWave). Low-band (sub-1 GHz) frequencies, such as 600 MHz (n71) in the US, 700 MHz (n28) worldwide, and 800 MHz (n20) in Europe, offer lower data speeds than higher bands, but are perfect for indoor and remote areas because they have long-range coverage and good penetration through walls. Mid-band (1-6 GHz) frequencies, such as 2.5 GHz (n41), 3.5 GHz (n78), 3.7-3.98 GHz (n77), and 4.9 GHz (n79), are used to balance speed and coverage. These bands are frequently used for urban and suburban deployments and offer speeds ranging from hundreds of Mbps to more than 1 Gbps. High-band (mmWave $>$ 24 GHz) frequencies, such as 26 GHz (n258), 28 GHz (n257), and 39 GHz (n260) offer extremely low latency and ultra-fast data rates of 1-10+ Gbps. Although mmWave bands have a limited range and poor penetration into buildings, they are suitable for applications where LoS connectivity is practical, such as in stadiums and crowded urban environments.

\paragraph{UAV Command and Control}
In India, the 5.825 and 5.875 GHz spectrum is used for UAV command and control. In the United States,  a dedicated frequency band 5030-5091 MHz is used for UAV command and control. This band offers greater reliability and less interference compared to the unlicensed 2.4 and 5.8 GHz bands.

%% file: userAssociation.tex
\section{User Association Problem}
\label{user association}

\begin{figure}[H]
    \centering    
    \includegraphics[width=2.5in]{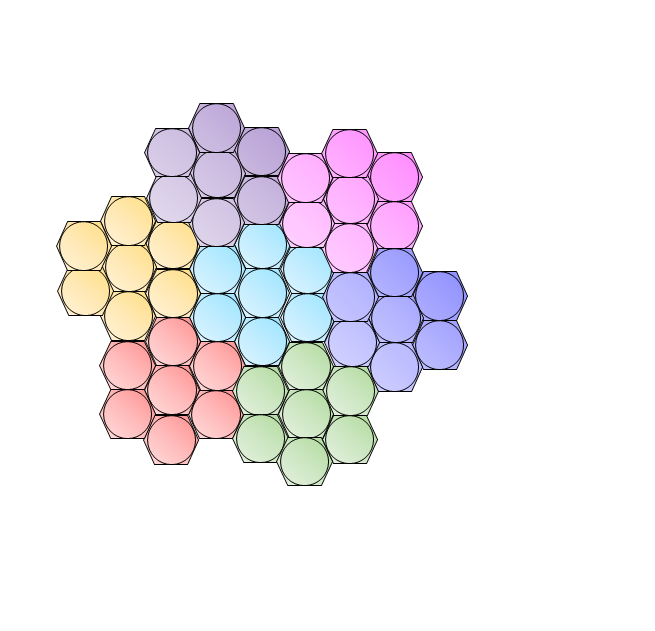}
     \caption{Cellular coverage area consisting of seven clusters each with seven cells.}
    \label{fig:coverage}
\end{figure}

\begin{figure}[H]
    \centering    
    \includegraphics[width=3in]{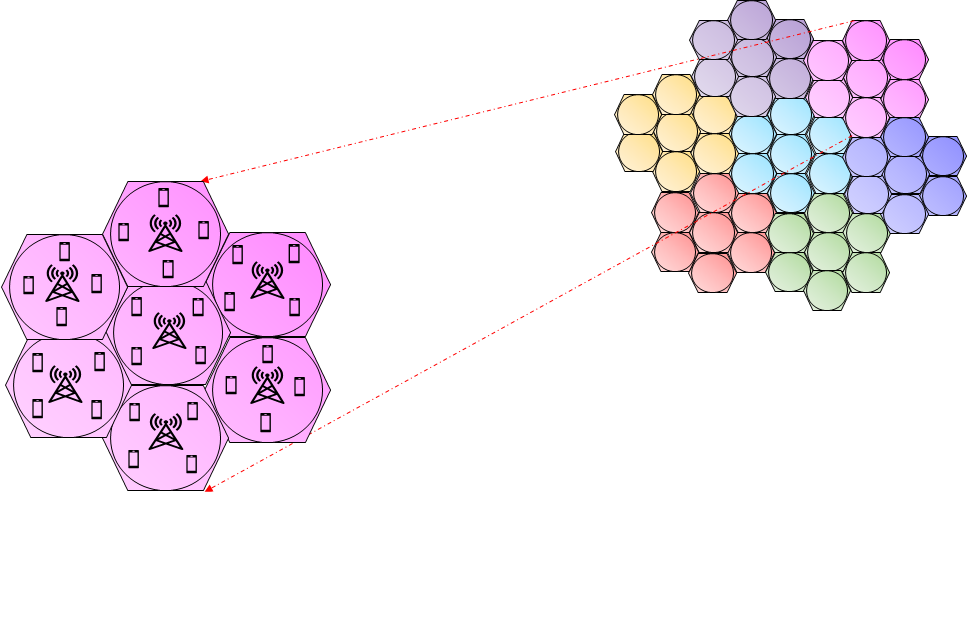}
     \caption{A cluster of seven cells each with its own base station.}
    \label{fig:basestation}
\end{figure}

Users are associated to an air segment or a ground segment based on their spatial locations, latency demands, and service priorities. In order to facilitate user association, the operational region is divided into a structured grid. Such a structured grid enables systematic coverage and allocation of resources. One such grid with seven clusters, each with seven cells, is illustrated in Figure \ref{fig:coverage}. Each cell has a fixed center and radius of coverage and includes a base station at its center, as shown in Figure \ref{fig:basestation}.

Under normal operating conditions, all users within a cell are associated with the base station located within the cell. However, when the base station is overloaded or becomes dysfunctional, some or all users need to be offloaded to the ABS that is temporarily deployed to accommodate them. The process of offloading users to one or more ABSs is known as user association. For example,  users with tight (lower) latency demand may be assigned to the TBS. On the other hand, users who can tolerate higher latency can be assigned to ABS if they are within its coverage range.  

%% file: mobility.tex
\section{Mobility}
\label{mobility}
This section outlines the mobility aspects in SAGIN. Mobility is present in all three segments of SAGIN. However, in this paper, user mobility and UAV mobility are studied in detail. Satellite mobility, in itself, requires an in-depth study and is therefore not discussed here. Whereas user mobility is natural, UAV mobility is explicitly designed to meet the coverage needs of the geographic region. 

\subsection{User Mobility}
User mobility refers to the movement of users within the service area of the network. During emergency situations, people move around for safety concerns, leading to service disruptions. User mobility introduces fading and interference, leading to variations in signal quality and increasing the risk of handover failures that may result in dropped connections.

\subsection{UAV Mobility}
\begin{figure}[h!]
    \centering    
    \includegraphics[width=3in]{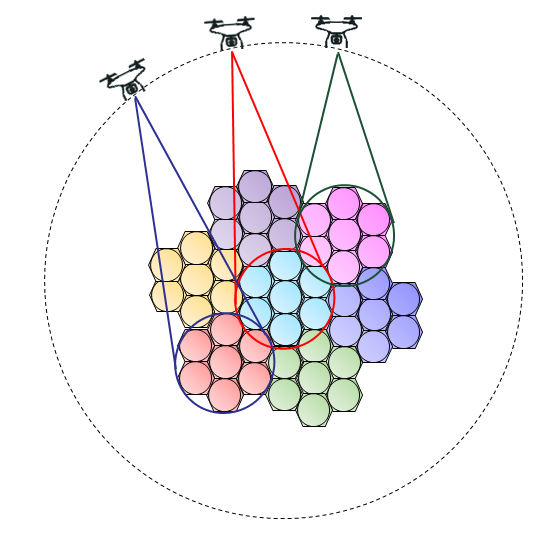}
     \caption{Mobility of three UAVs covering each cluster at a time} 
    \label{fig:UAVMobility}
\end{figure}

Figure. \ref{fig:UAVMobility} illustrates how UAVs move around to provide coverage to users on the ground. To ensure efficient and full coverage of all ground-based clusters, the circular coverage region is logically divided into three equal angular sectors, each covering 120 degrees. This division allows for a systematic deployment of UAVs in a rotational coverage strategy. In this example, three UAVs are deployed, each of which is responsible for covering a single group of base stations within a $120^\circ$ sector as shown in Figure 5. Each UAV's coverage is represented by colored triangular cones (blue, red, and green) that extend from the UAV down to its target cluster, while the clusters are represented by various color-coded zones made up of hexagonal cells. Once these three UAVs have completed coverage in their current sectors, they are collectively repositioned in the next $120^\circ$ sector. Every UAV in this new orientation supports a distinct cluster within the new sector. This sequential and rotational movement continues until all base station clusters have been served. This approach enables time-multiplexed UAV deployment throughout the ground network and significantly reduces the number of UAVs required to obtain complete area coverage. Furthermore, the approach facilitates dynamic reassignment of UAV tasks according to traffic load between clusters, latency tolerance, and demand. The dashed circle depicts the general footprint of an LEO satellite, and the UAVs work inside this line to provide the TBS with on-demand aerial support.

\subsection{Tradeoff between Mobility and Latency}
Mobility compensates for the limited availability of UAVs. If a cell becomes dysfunctional, a UAV equipped with a base station can replace the damaged site. In extreme scenarios where several towers are out of service, multiple UAVs, one for each affected cluster, may be required. When sufficient UAVs are not available, a smaller number can be deployed, and each UAV can move across clusters to provide intermittent coverage. In such cases, the UAV trajectories are determined by the coverage requirements. However, mobility introduces additional latency since users must wait for ABS to return to their cluster after moving out of range. In figure~\ref{fig:UAVMobility}, three UAVs are providing coverage for seven clusters.

\subsection{Satellite as a Backhaul}
\begin{figure}[h!]
    \centering    
    \includegraphics[width=4in]{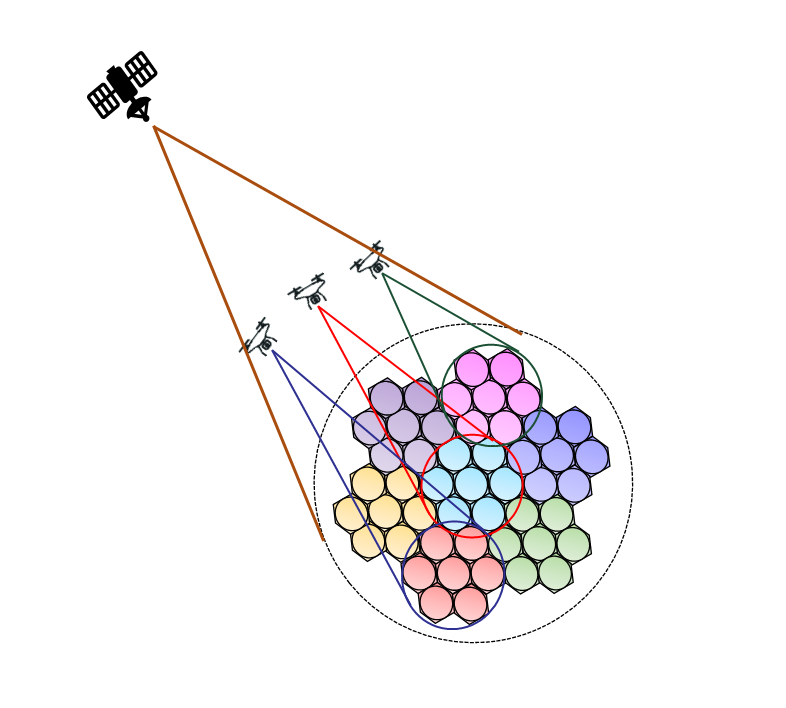}
     \caption{SAGIN with three mobile ABSs providing coverage for seven clusters each with seven cells, satellite providing a backhaul connectivity to the core network}
    \label{fig:SAGIN-3UAVS}
\end{figure}

In its simple form of SAGIN, shown in Figure \ref{fig:SAGIN-3UAVS}, the satellite provides a backhaul connection between an ABS and the core network. In this setup, the satellite acts as the relay node, ensuring that even when terrestrial infrastructure is unavailable or limited, the ABS can still maintain connectivity with the core communication network. More often, information may need to be exchanged among the three layers of SAGIN. For  example, UAVs can collect ground user data and relay it to LEO satellites \cite{chen2024intelligent}.   

%% file: resourceAllocation.tex
\section{Resource Allocation Problem}
\label{resource allocation}

\begin{table}[h]
\centering
\caption{Glossary of symbols used in SAGIN resource allocation model}
\begin{tabular}{|c|>{\centering\arraybackslash}p{5cm}|}
\hline
\textbf{Symbol} & \textbf{Description} \\
\hline
\( N\) & Total number of OFDMA subcarriers \\
\hline
\( K\) & Total number of users \\
\hline
\( R_{K,N} \) & Rate matrix \\
\hline
\( A_{K,N} \) & Resource allocation matrix \\
\hline
\( P_{K,N} \) & Power allocation matrix \\
\hline
\( n\) & OFDMA subcarrier index \\
\hline
\( k\) & user index \\
\hline
\( a_{k,n} \) & Element (k,n) of resource allocation matrix \\
\hline
\( p_{k,n} \) & Power allocated to user k on subcarrier n \\
\hline
\( r_{k,n} \) & Rate for user k on subcarrier n \\
\hline
\( \gamma_{k,n} \) & SINR \\
\hline
\( h_{k,n} \) & Channel gain for user k on subcarrier n \\
\hline
$\sigma^{2}$ & Noise power \\
\hline
\( B \) & Bandwidth of each subcarrier \\
\hline
\( P_{ABS}^{max} \) & Maximum power for ABS \\
\hline
\( r_{k}^{min} \) & Minimum rate for user k \\
\hline
\end{tabular}
\end{table}

When a TBS fails, users in the affected cell can be temporarily served by an ABS. If the ABS uses OFDMA, an effective resource allocation method is necessary to maintain service quality. The resource allocation problem involves the joint assignment of subcarriers and transmission power to users in a way that optimizes network performance while considering system constraints.

\subsection{System Model}

This section outlines the system model for the scenario one discussed in Section \ref{system architecture}. In this scenario, an ABS substitutes for a dysfunctional TBS. The OFDMA-based communication system is assumed to be used. Thus, the system consists of $K$ number of users and  $N$ number of OFDMA subcarriers. The resource block is represented in the matrix form given by \( A_{K,N} \) of $K$ rows and $N$ columns, the power allocated is represented by \( P_{K,N} \), and the rate is represented by \( R_{K,N} \). Throughout this paper, the three matrices, $A$, $K$, and $R$ are indexed through two indices, $k=1, \ldots, K$ and $n=1,\ldots, N$, where $k$ and $n$ represent the user and subcarrier respectively.

\subsection{Problem Formulation}
The objective is to maximize the sum rate for all $K$ users by optimizing the power allocation and subcarrier assignment. 
There are two sets of decision variables in this optimization problem, which are: 

\begin{equation}
\label{eq:a(kn)}
a_{k,n} \in \{0,1\},
\end{equation}
ia binary indicator, where $a_{k,n} = 1$, if the subcarrier $n$ is assigned to the user $k$, and $0$ otherwise, and

\begin{equation}
p_{k,n} \geq 0,  
\end{equation}
is the power allocated to user $k$ on the subcarrier $n$. 
The channel gain experienced by the user $k$ on the subcarrier $n$ is denoted by $h_{k,n}$, which depends on path loss, fading, and ABS altitude. The SINR,$\gamma_{k,n}$ for user $k$ on subcarrier $n$ is given by

\begin{equation}
\gamma_{k,n} = \frac{p_{k,n} h_{k,n}}{\sum_{j \ne k} a_{j,n} p_{j,n} h_{j,n} + \sigma^2} 
\end{equation}

\noindent The achievable rate ($r_{k,n}$) for user $k$ in subcarrier $n$, based on the Shannon-Hartley capacity formula, is 
\begin{equation}
r_{k,n} = B \cdot \log_2(1 + \gamma_{k,n})  
\end{equation}
where, $B$ is the bandwidth of each sub-carrier and ${\sigma^2}$ is the power of the noise. The objective is to maximize the sum data rate ($Sum~Rate$) of all users served by the ABS:

\begin{equation}
\max \limits_{a_{k,n}, p_{k,n}}  \sum_{k=1}^{K} \sum_{n=1}^{N} a_{k,n} r_{k,n}  
\end{equation}
subject to the following constraints:

\begin{enumerate}

\item Subcarrier exclusivity: Each subcarrier is allocated to at most one user.

\begin{equation}
\sum_{k=1}^{K} a_{k,n} \leq 1,\quad \forall n 
\end{equation}

\item ABS power constraint: The total power allocated must not exceed the ABS's maximum power budget.

\begin{equation}
\sum_{k=1}^{K} \sum_{n=1}^{N} a_{k,n} p_{k,n} \leq P_{\text{ABS}}^{\max} 
\end{equation}

\item User QoS constraint: A minimum data rate requirement $r_{k}^{\min}$ can be imposed for a guaranteed quality of service.

\begin{equation}
\sum_{n=1}^{N} a_{k,n} r_{k,n} \geq r_{k}^{\min}, \quad \forall k
\end{equation}

\item Binary subcarrier assignment: This is same as (\ref{eq:a(kn)}), but also listed here as a constraint.

\begin{equation}
 a_{k,n} \in \{0,1\}, \quad \forall k,n  
\end{equation}
\end{enumerate}

This optimization is a MINLP problem due to binary subcarrier assignment and non-linear SINR expression. It can be addressed using approaches such as the AO, DIWF, and GA, which are discussed below.

\subsection{Strategy 1: AO Algorithm}
\textit{1) AO algorithm with the same number of users and subcarriers}: The joint subcarrier and power allocation problem in a wireless communication system with an equal number of OFDMA subcarriers and users is solved using an AO algorithm  AO strategy was used previously for resource allocation in SAGIN  by other researchers \cite{nguyen2022computation},\cite{nguyen2023integrated},\cite{nguyen2022joint}. In this paper, AO is applied for subcarrier assignment and power allocations. The objective is to optimize the sum rate (total system throughput) by strategically assigning subcarriers to users and allocating the available transmit power among these subcarriers. Because of the mixed-integer nonlinearity of the problem, which consists of discrete subcarrier assignment variables and continuous power allocation variables, the optimization is split into two interdependent subproblems. First,  a non-linear programming solution (like \textit{fmincon ()} in MATLAB) is used to optimize the power allocation under a total power constraint, assuming a fixed subcarrier assignment. Then each subcarrier is assigned to the user who gets the maximum data rate on it, updating the subcarrier assignment greedily while keeping the power fixed. These two steps are repeated alternately until convergence is achieved, typically when the assignment variation between iterations becomes insignificant. This algorithm provides a practical and computationally efficient solution for resource allocation problems in a dynamic wireless environment like SAGIN.

\textit{2) AO with more users than subcarriers}: When the number of users exceeds the available subcarriers, resource allocation must balance efficiency and fairness. Since each subcarrier can serve only one user at a time, time-domain scheduling is used, where different user groups share subcarriers across slots. User selection depends on channel conditions, QoS requirements, and fairness metrics such as round-robin or proportional fairness. In emergency communication scenarios, priority-based allocation is critical. Using an AO algorithm, subcarriers and power are dynamically assigned. First responders ( firefighters, police, medical teams) receive the first slot to ensure low latency and high reliability, while later slots redistribute resources among civilians for equitable access. This alternate scheme provides mission-critical service to responders while maintaining civilian connectivity and adapting dynamically to mobility, emergencies, and real-time channel conditions without requiring extra spectrum.

\subsection{Strategy 2: Subcarrier Assignment followed by DIWF for Power Allocation}

This method distributes power among users in an OFDMA system, while ensuring that every user reaches the minimum required data rate. Of the $N$ possible subcarriers, a unique subcarrier is assigned to each of the $K$ users. Initially, the algorithm uses Shannon's capacity formula to convert the specified minimum rate limitation into a minimum SNR, which determines the minimum power needed for each user to meet it \cite{dai2020power}. The problem is considered infeasible if the sum of these minimal power requirements is greater than the available power budget. Otherwise, the remaining power is distributed iteratively.

Unlike the conventional water-filling approach, DIWF introduces a damping factor that combines the previous allocation with the current water-filling update. This prevents oscillations, enforces monotonic improvement, and ensures smooth convergence toward the optimal solution. At each iteration, the total achievable sum rate and rate per user are recalculated based on the updated power allocations. The process continues until the change in sum rate between iterations falls below a predefined tolerance.

In this way, the DIWF algorithm balances efficiency and robustness. Stronger channels are favored in the water filling stage to maximize spectral efficiency, while weaker users are guaranteed sufficient power to meet their minimum rate constraints. The damping mechanism ensures that the convergence is stable even under random channel variations. 


\subsection{Strategy 3: GA}

GA based method is proposed here for joint subcarrier assignment and power allocation. This algorithm seeks to optimize the sum rate  \cite{hammami2025meeting} while adhering to the constraints of subcarrier exclusivity, total transmit power, and per-user QoS. The user-subcarrier mapping and related power allocation are encoded by every individual of the population. Power allocations are optimized within a limited range and subcarrier mappings are represented using permutation matrices to guarantee exclusivity. A fitness function assesses each solution according to QoS, power budget, and acceptable data rate. To evolve the population, standard GA operations including selection, crossover, mutation, and repair are performed iteratively. The best solution across generations provides the optimal resource allocation configuration.

%% file: simulations.tex
\section{Simulations}
\label{simulations}

\subsection{Simulation Environment}
\begin{table}[H]
\centering
\caption{Common parameters used in all simulations}
\label{parameters AO}
\begin{tabular}{|c|>{\centering\arraybackslash}p{3cm}|}
\hline
\textbf{Symbol} & \textbf{Description} \\
\hline
\( N\) & 12 \\
\hline
\( K\) & 12 \\
\hline
 $\sigma^{2}$ & 10{$^{-9}$} \\
\hline
\( B \) & 100 KHz \\
\hline
\( P_{ABS}^{max} \) & 20 Watts \\
\hline
\( max\ iterations \) & 100 \\
\hline
\(MC\ runs\) & 10000 \\
\hline
\end{tabular}
\end{table}

In this section, to evaluate the performance of the proposed resource allocation strategy, the results were simulated using MATLAB. Three different optimization algorithms were tested and implemented: alternating optimization, water filling, and genetic. For each algorithm, 10,000 Monte Carlo simulation runs were performed, modeled channel gains as stochastic variables to capture the impact of channel randomness, ensuring statistically reliable results.

\subsection{Strategy 1: Alternating Optimization}
\begin{figure}[H]
    \centering    
    \includegraphics[width=3.4in]{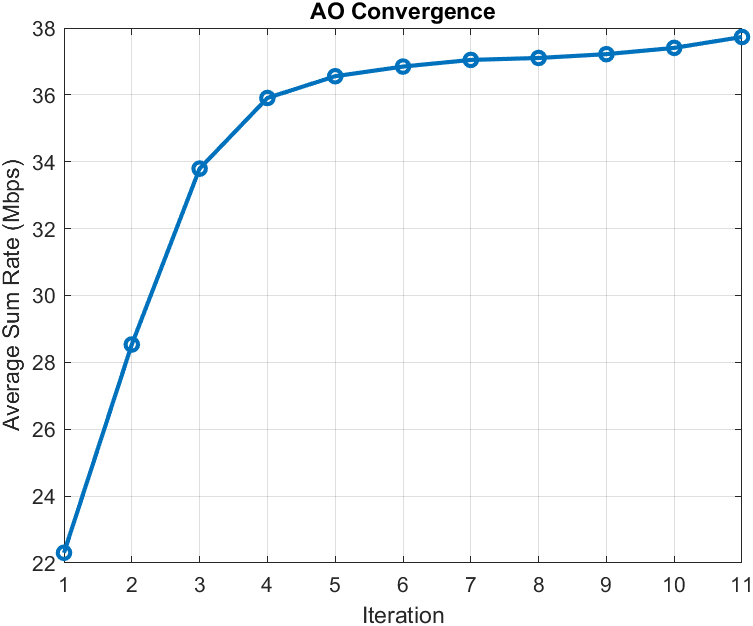}
     \caption{Monte Carlo average sum rate of AO algorithm convergence.}
    \label{fig:MonteCarloAO}
\end{figure}

Figure \ref{fig:MonteCarloAO} shows the Monte Carlo-averaged convergence behavior of the AO algorithm. The horizontal axis represents the number of iterations, while the vertical axis represents the average sum rate in Mbps. For each realization, the algorithm was initialized with random subcarrier assignments and power allocations, and then iteratively updated these variables in an alternating fashion.

The channel gains are randomly generated in every MC trial; therefore, the exact number of iterations required for convergence varies from run to run. However, in every individual run, the AO algorithm guarantees a monotonic increase in the sum rate with respect to the iteration index, since each alternating step optimizes the sub problem. The figure depicts the average trajectory of these monotonic improvements in all trials.

In the simulations, the runs converged between 2 and 11 iterations. This variability explains why the number of runs that contribute to the trials decreases as the iteration index increases. For example, in iteration 3, 9994 trials were active, while in iteration 7, only 1713 trials continued. Beyond iteration 10, only a handful of runs remained, which the averaging process accounts for. At each iteration index, the average sum rate is computed only for the subset of trials that reached that iteration. This approach ensures that the curve represents the progression of the AO algorithm while respecting the fact that not all runs last the same number of iterations.

From the average curve, it can be observed that the performance gain occurs during the initial iterations. The average sum rate increases from iteration 1 to iteration 3. Beyond this point, the improvements become more gradual, and the curve begins to flatten out. Around iteration 10, the average sum rate stabilizes, meaning that the AO algorithm has converged under the simulation conditions considered. The plot shown in Fig. \ref{fig:MonteCarloAO} appears to benefit from more iterations, i.e, the sum rate might increase if the number of iterations are increased beyond 11, but it is not the case. For sum rate to increase, the subcarrier assignment needs to be improved. However, the subcarrier assignment stabilized within 11 iterations during the 10,000 Monte Carlo simulations, which led to the convergence of the sum rate within 11 iterations. The exact number of iterations will, of course,  depend on the specific channel conditions among other settings.

\subsection{Strategy 2:Damped Iterative Water Filling Algorithm}
\begin{figure}[H]
    \centering
    \includegraphics[width=3.4 in]{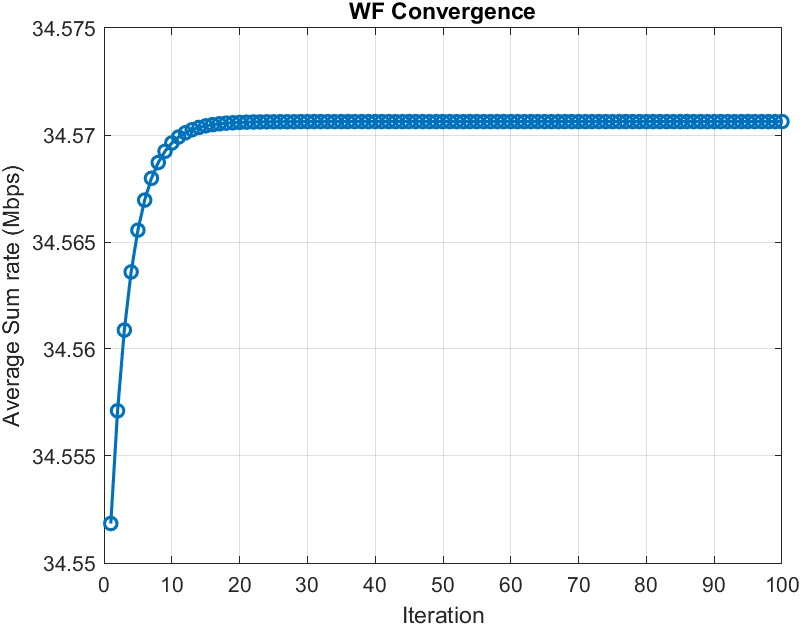}
    \caption{Monte Carlo average sum rate of  damped iterative WF algorithm convergence.}
    \label{fig:WF}
\end{figure}

\begin{table}[H]
\centering
\caption{Parameters specific to DIWF algorithm simulation.}
\label{parameters GA}
\begin{tabular}{|c|>{\centering\arraybackslash}p{3cm}|}
\hline
\textbf{Symbol} & \textbf{Description} \\
\hline
\( R_{min}\) & 0.1 \\
\hline
\( Damping \ factor (\alpha) \) & 0.15 \\
\hline
\( tolerance \) & $10^{-5}$\\
\hline
\end{tabular}
\end{table}
Figure \ref{fig:WF} represents the convergence of the DIWF algorithm for power allocation under one-to-one subcarrier-user mapping. The process begins with an initial allocation and iteratively updates the power distribution using a damping factor that blends the previous allocation with the optimal water-filling solution. This modification ensures smooth convergence and prevent oscillations, which occur in conventional iterative WF when channel randomness or QoS constraints are present.

The convergence plot shows that the DIWF algorithm steadily improves the system sum rate. Within the first few iterations, the sum rate rises rapidly, after which the improvement becomes gradual and the curve stabilizes, indicating convergence. The monotonic increase confirms that each update moves the allocation closer to the optimal solution without overshooting.

The use of user-specific minimum power guarantees ensures that weaker channels are allocated sufficient power to meet their minimum power requirements, whereas users with strong channel gains benefit from a larger share of the remaining power. Monte Carlo averaging further ensures that the curve reflects reliable performance under random channel variations. Overall, the results validate that DIWF provides a robust and stable graph by achieving the optimal performance while maintaining feasibility and convergence. 

\subsection{Strategy 3: Genetic Algorithm}
\begin{table}[H]
\centering
\caption{Parameters specific to genetic algorithm simulation.}
\label{parameters GA}
\begin{tabular}{|c|>{\centering\arraybackslash}p{3cm}|}
\hline
\textbf{Symbol} & \textbf{Description} \\
\hline
\( Population\ size \) & 40 \\
\hline
\( Number\ of\ generations \) & 30 \\
\hline
\( Mutation\ rate \) & 0.1 \\
\hline
\end{tabular}
\end{table}

\begin{figure}[H]
    \centering    
    \includegraphics[width=3.4in]{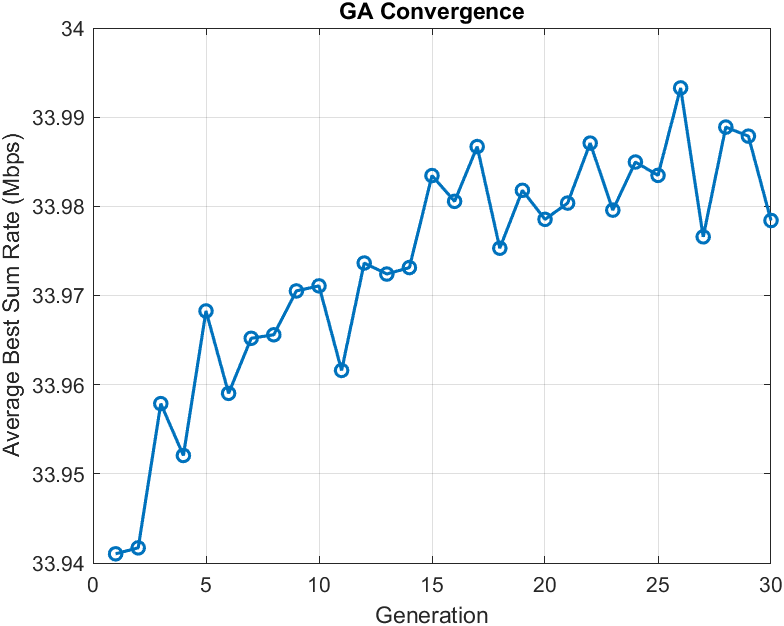}
     \caption{Monte Carlo average sum rate of genetic algorithm convergence.}
    \label{Monte Carlo simulation of Genetic Algorithm}
\end{figure}

Figure~\ref{Monte Carlo simulation of Genetic Algorithm} shows the convergence behavior of the GA over 30 generations. The average best sum rate increases steadily from about 33.94 to nearly 34 Mbps, indicating consistent evolutionary improvements. Although the curve exhibits small fluctuations due to crossover and mutation, the overall trend demonstrates that the GA is able to refine the population towards higher-performing solutions across generations.

The optimization process yields a best fitness (sum rate) of 36.55 Mbps, while the average across 10,000 Monte Carlo runs is 35.04 Mbps, highlighting the robustness of the GA in different random channel realizations. The best subcarrier assignment matrix ensures that each user is allocated a distinct subcarrier, while the corresponding power allocation matrix distributes the available 20 Watts budget in a non-uniform way. This adaptive allocation reflects channel conditions, assigning more power to users with favorable gains and less to those with weaker channels, thereby maximizing the overall throughput while meeting constraints.

%% file: conclusions.tex
\section{Summary, insights, and conclusions}
\label{concl}
This paper investigated SAGIN with a focus on air and ground segments. Resource allocation and user association strategies are studied in detail. Resource allocation is studied in the context of an OFDMA-based communication system, where user association is studied in terms of ABS and TBS.

\subsection{Summary} In summary, the main contributions of this paper included: (1) a thorough study and analysis of three resource allocation algorithms, AO, DIWF, and GA, in terms of their convergence, and (2) user association strategies in terms of their applications during emergency situations.

\paragraph{AO Algorithm} The AO algorithm jointly optimizes power allocation and subcarrier assignment by iteratively updating them until convergence. Power is allocated using convex optimization, while subcarriers are greedily assigned on the basis of achievable rate. Monte Carlo simulations were used to reduce the randomness of channel gains and ensure reliable performance evaluation. The sum rate improves over iterations and stabilizes, demonstrating that the AO algorithm effectively balances user-subcarrier mapping and power distribution under system constraints.

\paragraph{DIWF Algorithm} The DIWF algorithm is used to achieve robust power allocation under one-to-one subcarrier and user mapping. Unlike conventional one-shot water filling, DIWF updates power allocations gradually using a damping factor, which avoids large fluctuations and ensures stable convergence. User-specific minimum power levels are enforced to satisfy minimum rate constraints, while Monte Carlo simulations are performed to average out channel randomness and ensure reliable performance evaluation. The algorithm converges to the optimal water filling target while maintaining the total power budget, and the sum rate versus iteration plots confirm stable and efficient convergence.

\paragraph{GA} A GA is applied to jointly optimize subcarrier assignment and power allocation. Each chromosome encodes both the assignment matrix and power allocation, while feasibility is ensured through a repair function that enforces a one-to-one user and subcarrier mapping. The fitness function maximizes the total achievable rate while imposing penalties for violations of the total power budget and minimum rate requirements. It's operators including roulette wheel selection, single point crossover and mutation on the power allocation part, drive the population toward improved solutions across generations. The results demonstrate that GA effectively converges to a feasible allocation strategy that balances throughput, fairness, and system constraints.

\subsection{Insights} Several insights gained during this research study are discussed in the following that explain the problem and the results of the solution strategies. 


\paragraph{DIWF}  DIWF is fundamentally a closed-form one-step solution when the problem is convex and involves a single user with known channel gains. However, when moving into multi-user or networked resource allocation settings, the problem quickly becomes more complex due to interference coupling, fairness constraints, or joint optimization across heterogeneous resources (time, frequency, space, and power). In these cases, DIWF is often adapted into iterative schemes.

\paragraph{Nature of AO} AO studied in this paper belongs to the general class of optimization methods known as Alternating Direction Method of Multipliers (ADMM) which decomposes a global optimization problem into smaller subproblems that can be solved in parallel, with a coordination mechanism ensuring consensus. Iterative water filling can be viewed as a primal update step within an ADMM framework, where each user (or subcarrier) locally updates its power allocation given the current subcarrier assignment. The size of the steps in iterative water filling parallels the penalty parameter in ADMM, which controls convergence behavior. ADMM provides theoretical guarantees on convergence under convexity, making it attractive for SAGINs.

\paragraph{Convergence behavior of GA} GA did not exhibit smooth convergence for several reasons related to its design principles. Unlike gradient-based methods, which iteratively refine a single solution, GA works with a population of candidate solutions. At each generation, selection, crossover, and mutation can introduce large jumps in solution quality, causing convergence to look noisy or erratic rather than smooth.

\subsection{Conclusions}

The paper addresses the resource allocation and user association challenges in SAGIN by evaluating three key algorithms, AO, DIWF and GA under OFDMA communication framework for UAV based ABSs. Through extensive simulation- based performance analysis, it is demonstrated that each method offers distinct advantages. AO achieves rapid convergence and balanced throughput, DIWF provides robust power allocation with guaranteed user rate constraints, and GA delivers flexible, near-optimal solutions that benefit from population diversity. Implementing these strategies achieves notable improvements in throughput, fairness, and latency, confirming the effectiveness of combining mathematical optimization and learning-based resource management techniques.



%% file: ack.tex
\section*{acknowledgments}
The research presented in this paper is based upon work supported by the National Science Foundation under grants CNS -2148178 (Resilient \& Intelligent NextG Systems) and CNS-2330773 (Building a Smart and Connected Ecosystem for the First-Responder Community). The authors also would like to thank Skyler Hawkins for carefully reviewing this paper and making suggestions for improving the quality of presentation.